# An empirical evaluation for defining a mid-air gesture dictionary for web-based interaction

Thomas Pasquale
Department Name
University of Turin, Italy
thomas.pasquale@unito.it

Cristina Gena
Computer Science Department
University of Turin, Italy
cristina.gena@unito.it

Fabiana Vernero
Computer Science Department
University of Turin, Italy
fabiana.vernero@unito.it

## ABSTRACT

This paper presents an empirical evaluation of mid-air gestures in a web setting. Fifty-six (56) subjects, all of them HCI students, were divided into 16 groups and involved as designers. Each group worked separately with the same requirements. Firstly, designers identified the main actions required for a web-based interaction with a university classroom search service. Secondly, they proposed a set of mid-air gestures to carry out the identified actions: 99 different mid-air gestures for 16 different web actions were produced in total. Then, designers validated their proposals involving external subjects, namely 248 users in total. Finally, we analyzed their results and identified the most recurring or intuitive gestures as well as the potential criticalities associated with their proposals. Hence, we defined a mid-air gesture dictionary that contains, according to our analysis, the most suitable gestures for each identified web action. Our results suggest that most people tend to replicate gestures used in touch-based and mouse-based interfaces also in touchless interactions, ignoring the fact that they can be problematic due to the different distance between the user and the device in each interaction context.

## 1 Introduction

Touchless interaction is becoming a popular way to interact with devices, with no need for physical contact. This is especially true in public spaces, where touchless interaction can be faster than using a mouse or a keyboard, also due to the frequent use of large displays [2]. According to Ardito et al. [2], hand gestures are those whose interaction modality is through remote gestures (namely performed by the user without any contact with the display, thus touchless), and they are also called *mid-air* or simply *air* gestures. In this paper, we will refer to these gestures and we will call them *mid-air gestures*.

Web browsing is one of the most common tasks that users perform on their devices. However, touchless web browsing can be challenging, as it can be difficult to move a cursor or carry out other tasks without traditional input devices.

This paper presents an empirical analysis of mid-air gestures for touchless interaction with a web application. The study involved a sample of 56 participants, divided into 16 groups, who designed a set of mid-air gestures for web-based interaction and, specifically, for interacting with a university classroom search service. Then, they validated their proposal involving external subjects, namely 248 users in total, with an average of 15.5 users per group.

Based on their proposals, we devised a final dictionary which includes the most suitable mid-air gestures and is optimized to guarantee a fluid web interaction. Our results provide valuable insights for the design and improvement of future mid-air gesture-based interactions.

This paper has been organized as follows: Section 2 briefly discusses related works, focusing on contexts for the application of touchless interaction and proposals for the design of distinctive gestures. Section 3 presents the methodology we followed for our empirical study, while Section 4 discusses our results, presenting the web actions included in our dictionary and the gestures we selected for each of them. Finally, Section 4 concludes the paper by discussing contributions, limitations and future directions for this research.

## 2 Related work

*Mid-air gestures* call for natural interfaces [9], which refer to user interfaces that are invisible, or become invisible with successive learned interactions, to their users, and have been applied in several different domains. While the focus in this paper is primarily on web-related tasks, other studies address the alphabet's representation through gestures, without using an on-screen keyboard [11]. In addition, touchless interaction can help users with disabilities in interacting with devices, for example by translating sign language into text, thus easing communication with other people [11]. On a different note, touchless interaction can also be used in shops to attract customers through innovative forms of interaction and more engaging forms of advertising [5]. The use of large displays in combination with touchless gestural interaction has also been proposed in the context of the Smart Industry [7]. For example, in [1] the project consortium engineered a smart armband able to detect gestures through the analysis of both movement and muscle bio signals, while a machine learning library allows to calibrate and recognize task-specific gestures. The definition of an appropriate set of gestures undergone several steps, including a guessability study.

To interact with a touchless device, gestures must be identified and translated into commands understandable by the system. The use of colored caps worn on fingertips has been proposed to make the identification of gestures corresponding to cursor actions easier [12]. In this scenario, the left click is produced by moving the index and middle finger close to each other, while the index finger (i.e., with the detection of a single-color cap) is considered enough to perform a right click [12]. Other studies [9] propose the use of the index and middle finger for the double-click confirmation, or a pinch using thumb and index fingers for drag and drop tasks, while, to control the cursor, the index finger movement is mapped to the representation of the cursor.

System activation can be accomplished in different ways, such as holding the palm open, using a closed fist, making a peace sign, transitioning from a fist to an open palm, or a combination of finger gestures [10].

Instead of proposing a wide range of actions to the user, one way is to limit the actions to elementary ones, such as using the palm for stopping and a thumbs-up gesture for confirmation. Then, through the user interface (UI), the user is guided to successive choices, further simplifying the interaction but limiting the user's options, as for instance in [3].

Instaed of deploying an exhaustive on-screen keyboard interface for textual input, an alternative modality entails the presentation of a miniature virtual keyboard, preconfigured with indispensable typing keys. This pre-mapped keyboard is is located on the top of the screen in a red box. Users engage with the system by horizontally sweeping their open palm across the preconfigured keyboard, thereby accessing the alphabet characters and keyboard functions. To select a specific key, the users place their finger over the chosen key [8].

## 3   The study

The goal of the study was to define a mid-air gesture dictionary for a web-based interaction, tested in the context of a university classroom search service. To this aim, we analyzed a large set of mid-air gestures proposed and validated by a large sample of users, to empirically identify the most prevalent, intuitive, and shared ones.

The study originated in a university setting, during an assignment for a Human-Computer Interaction (HCI) master course. Users were involved on two occasions: firstly, students acted as designers to propose an initial set of mid-air gestures and, secondly, external users, not involved in the design phase, were engaged to collect external feedback. In addition, a few groups also involved external users as co-designers in the design phase.

Involved external users were asked to submit a consent form to take part in the research, while students as designers were asked permission for using their material in this research.

### 3.1 Participants

Fifty-six (56) students, 40 female (71%) and 16 male (29%) were involved *as designers*. They were divided into 16 groups, ranging in size from a minimum of 2 to a maximum of 4 members. Since designers were all master students, all of them had at least a bachelor's degree.

Five (5) out of 16 groups (31%) consulted *25 external co-designer*s (5 per group) to get help in finalizing their proposal.

All 16 groups submitted their proposed gestures for external review. The total number of external users involved in this phase was 248, with a modal value of 4 external users per group.

The overall number of external subjects involved in the experiment is 273, with age ranging from 18 to 55 years old, with a majority of females (68%). Their educational background varies from a high school diploma to a bachelor's degree. As for their technological expertise, both beginners and experts were recruited.

### 3.2 Apparatus and Materials

The groups of designers used personal devices, such as laptops, to simulate touchless devices to be used during the mid-air gesture generation phase.

Videos showing the designed gestures, to be used to collect external feedback, were captured using an external camera and recorded in .mp4 or .gif formats.

A web form embedding the aforementioned videos was used to collect structured feedback about the designed gestures.

### 3.3  Procedure

The study included two macro-phases: the generation of a large number of mid-air gestures and the definition of a dictionary optimized for web-based interaction. In the first phase, the following activities were carried out: identification of meaningful web actions, design of mid-air gestures and gesture validation. In the second phase, the main steps were the analysis of the identified actions and the analysis of the proposed gestures for each action.

*3.3.1 Mid-air gesture generation.* The 16 groups of designers were required to design a touchless interaction assuming that a large display was installed on the university campus. In particular, the screen would display the university website with schedules for classes and classroom information and students would be able to use a classroom search service, as shown in the prototype in Fig 1.

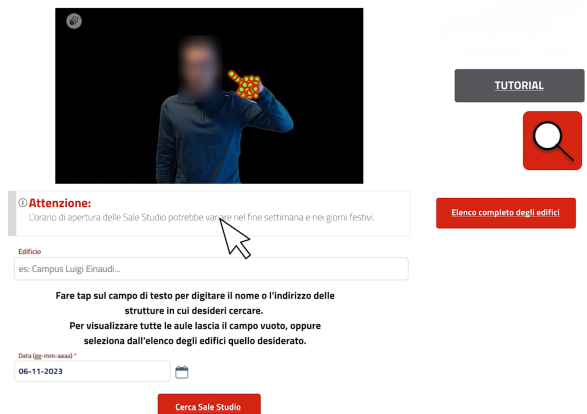

Figure 1. The web prototype for the touchless interaction envisioned by a group of students,

As a first step, each group had to identify the main actions required to carry out basic tasks in the aforementioned scenario. Examples of such actions are: moving the cursor, clicking confirmation and page resizing.

After that, each group had to come up with mid-air gestures that would enable users to carry out such actions, browsing the webpage effectively. Groups were suggested to reach out to external users for help and inspiration. Thus, 5 groups (31%) chose to consult 25 external co-designers (5 per group). Co-designers were asked to enact the gestures they would make on a touchless screen to perform the previously identified actions, and the most popular ones were chosen by the designers to be part of their proposal. The other groups worked autonomously in this phase.

Once each group had defined an initial set of gestures, they were required to have them reviewed by external users, to ensure they were easy to understand. To do so, a Google Form was used which included video representations of each gesture to evaluate. More specifically, external users had to associate each gesture to the most appropriate action, choosing from a list which included all the previously identified actions. External users could also provide suggestions on ways of improving gesture execution, either by providing free text comments or by submitting anonymous video proposals, if needed. Based on the feedback they received, the groups had the opportunity to tweak the initially defined gestures if necessary: in particular, 8 groups (50%) further modified their gestures to develop a definitive version incorporating the suggested changes.

Finally, each group was required to submit their final proposal to the authors of this paper. Submissions should be in the form of an explanatory video illustrating all the gestures, accompanied by a descriptive report.

*3.3.1 Dictionary definition.* Firstly, we analyzed the lists of actions proposed by each group, to identify commonalities and merge similar actions which were labeled differently by different groups. Secondly, we separately considered each action and examined all the devised mid-air gestures, trying to identify commonalities in the proposals of different groups. In both cases, we used categorization and counting [4], an approach to the transposition of quantitative data into qualitative data which implies the definition of recurring categories and can be considered a simplified version of thematic analysis [6]. Then, we closely examined the resulting categories, not only considering their popularity, but also trying to anticipate the experience of users performing the defined gestures. Hence, we identified the most intuitive, and suitable mid-air gestures as well as the potential criticalities associated with their application. Finally, following the results of our analysis and trying to maintain consistency between all the chosen options, we selected the most suitable mid-air gestures for each identified web action, thus defining a dictionary optimized for web-based interaction.

## 4      Results

The designers' groups proposed a set of mid-air gestures for various functionalities, including cursor-pointing, click confirmation, vertical and horizontal scrolling, page resizing, quick history navigation, panning in resized pages (zoom >100%), drag and drop, interaction activation and interaction ending, homepage access, page reload, close current page, context menu access, volume control, multiple items selection, and text selection. In total, 99 different version of these gestures were designed analyzed as part of this study.
However, not all gestures were included in the final dictionary as they were considered unnecessary for achieving optimal and fast interaction. The following gestures were excluded from the analysis: closing the current page, fast access to the context menu, volume control, multiple items selection, and text selection.

The gesture for closing the current web page was omitted from the analysis since in a touchless interaction with a public display, end users are unlikely to possess the authorization to close the current page. In situations where a page has been designed for public interaction, it is advisable to restrict users from closing it and exiting the touchless environment.
Gestures for accessing the context menu, volume control, multiple selections of elements, and text selection were excluded on the grounds of being non-essential for this scenario. Instead of the text selection gesture, consideration may be given to developing a gesture aimed at swiftly deleting characters within a text field.
In the followings, we will present the gestures we considered for our proposed dictionary.

### 4.1      Cursor pointing

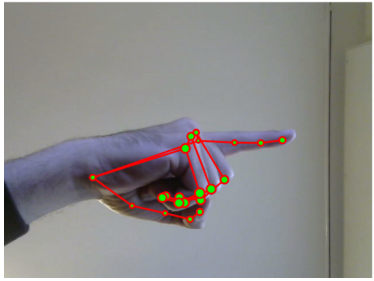

**Figure 2**: Cursor-pointing gesture.[1].

Out of the 16 groups, the majority, comprising 9 groups (56%), opted for the index finger as the virtual representation of the cursor (see Fig. 2). Meanwhile, 4 groups (25%) chose an open hand positioned vertically, synchronizing its movement with the cursor. 2 groups (13%) employed both the index and middle fingers for cursor movement, while one group (6%) utilized the index, middle, and thumb collectively to control the cursor.

The most detected gesture is the use of the index finger (Fig. 2). Although a valid alternative is the use of an open hand, both gestures are deemed acceptable. However, a preference is expressed for the use of the index finger, primarily because of its widespread use in conventional mouse interactions. This choice not only capitalizes on the familiarity of users with mouse-based interactions but also enhances the intuitiveness and seamless transition for individuals accustomed to traditional interfaces.

## 4.2 Click confirmation

8 groups out of 16 (50%) choose a quick movement of the index finger toward the webcam. 3 groups (19%) used a double-tap with the index finger in front of the display, while 2 other groups (13%) proposed the same gesture but with two fingers. 2 groups (13%) opted for closing the entire hand over the desired element. 1 group (6%) used the index, middle, and thumb joining together to complete the action. Building upon the pointing gesture, our investigation was focused on determining the optimal combination to trigger the confirmation action.

As previously mentioned, the groups conducted their studies with users interacting closely with the laptops. Given the potential deployment of the system on a kiosk, it is reasonable to assume that end users may not be in as close proximity as in the analyzed case but rather at a greater distance. This adjustment in user positioning should be considered when designing and evaluating the system for practical application at a greater distance.

The gesture most frequently observed entails a rapid movement of the index finger towards the display. Although a valid choice, it is essential to note that users may adopt an outstretched arm for pointing, and executing the gesture might be impractical without physically moving toward the device. Consequently, this gesture has been excluded from the final dictionary, taking into account practical concerns.

A more suitable alternative could be to employ the second most recurring gesture, namely, an index-finger double tap. Enhancements can be made to this gesture by introducing a timing element for lowering the finger, thereby eliminating the need for a double tap. This refinement aims to streamline the interaction process and improve the overall usability of the gesture.

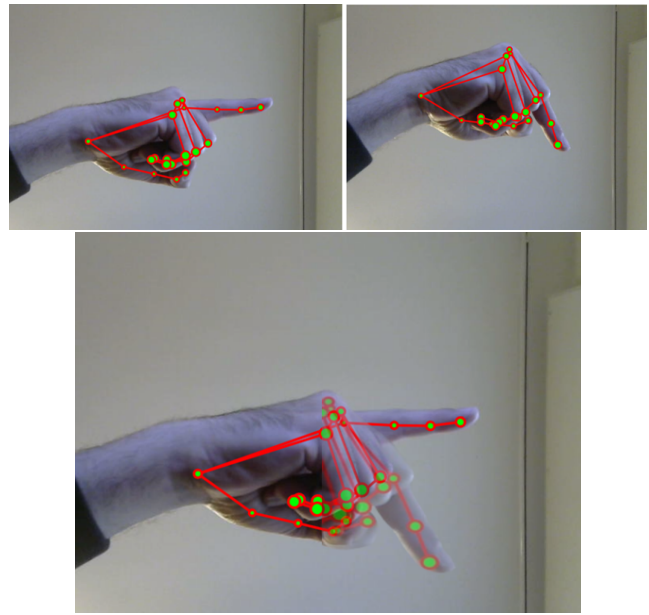

**Figure 3**: Confirmation gesture.

As depicted in Figure 3, this gesture is easily replicable in a touchless environment. Nevertheless, it is recommended to augment the range of the index finger movement for a more distinct and easily detectable gesture. Additionally, a minimum holding time of $t \geq 2s$ is introduced to execute the action. While this choice may lead to a slower interaction due to the mandatory wait time for users when confirming an action, it is believed that this approach will substantially decrease errors

---

[1] The hand landmarks were highlighted using Google MediaPipe Hand landmarks detection algorithm, https://developers.google.com/mediapipe/solutions/vision/hand_landmarker

resulting from misinterpretations, while still ensuring a more reliable and positive outcome.

## 4.3 Vertical and horizontal scrolling

One group (6%) did not create a gesture for this task. 8 out of the remaining 15 groups (54%) opted to use an open hand perpendicular to the display intending to move vertically or horizontally according to the desired direction. Among these, 3 groups (20%) also design an acceleration in scrolling if the hand movement is intensified. In contrast, all others (80%) presumed a movement directly proportional to the hand's motion. 3 groups (20%) chose to use index and middle fingers in a fixed horizontal position parallel to the webcam, moving in the desired direction. 2 groups (13%) considered a synchronized and repeated movement of pinky, ring, middle and index finger fixed together in the desired direction. One group (7%) used the index thumb and middle finger, which by joining together can proceed to move the page in the required direction.

The most popular gesture for vertical or horizontal scrolling involves the use of an open hand; however, this gesture is not considered for this action due to the current configuration of the dictionary. Until now, all actions have been performed using the index finger, and we aim to maintain consistency for fundamental gestures to be as agile and intuitive as possible. While the user is already moving the cursor through the index, it is desirable to add the middle finger to activate the scrolling functionality, configuring it as the definitive gesture, see Fig. 4.

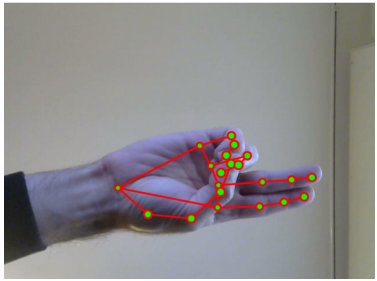

**Figure 4**: Scrolling gesture.

This action inevitably resembles the scrolling function on laptops, where placing two fingers on the touchpad starts the scrolling action. Consequently, this gesture may have a high guessability, thanks to the previous experience acquired on other devices.

## 4.4 Page resizing

The page resizing gesture for the page comprised two distinct actions: zoom in and zoom out. In each designer group the creation of the zoom-in gesture was consistently followed by the development of a contrasting gesture to facilitate zoom-out, ultimately restoring the page to its default size.

Out of the 16 groups studied, five groups (31%) opted for a gesture involving both hands initially placed adjacent to each other, followed by expanding them outward. Another group (6%) proposed a similar version, beginning with hands touching together. Four groups (25%) selected the opening of the index finger and thumb. Another 4 groups (25%) chose the simultaneous opening of all fingers to execute the zoom-in action. One group (6%) utilized the opening of the index, middle finger, and thumb for zooming in. A different group (6%) preferred starting with all fingers closed, gradually opening the hand while moving closer to the display. Lastly, one group (6%) employed both hands, joining the thumb, middle, and index fingers, with their expansion triggering the zoom-in action.

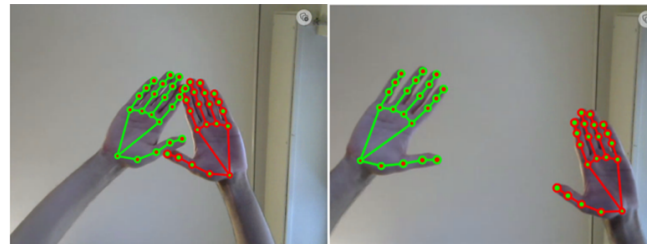

**Figure 5**: Page resizing gesture.

The most widely adopted gesture for resizing the page involves the expansion of both hands, as illustrated in Figure 5. Recognizing that the action of resizing the page is not as frequent as tasks like pointing or confirmation, the justification for employing a different hand/fingers configuration for this specific task is justified. Consequently, this distinctive gesture has been included in the final dictionary.

## 4.5 Quick browsing history

Three groups (19%) did not propose this gesture. From the remaining 13 groups, the majority, 9 groups (69%), opted to navigate in history by employing the entire hand to make a quick movement to the left. One group (8%) chose a rapid movement to the left using their hand, specifically with the thumb raised and the other fingers closed. Another group (8%) decided on a quick movement by shifting the right elbow to the left. Yet another group (8%) envisioned a clockwise circular movement of the entire right hand. Finally, one group (8%) selected using two fingers to swipe to the left.

Among those who created this gesture, 5 groups (38%) created a gesture exclusively to go back to the recently visited page. Meanwhile, 8 groups (61%) created a gesture for both backward and forward navigation in history. For these groups, the gesture for moving forward is identical to the one for moving backward, but the direction of the movement is reversed.

The most frequently recurring gesture for navigating in history involves using the entire hand to make a swift movement to the left or right, depending on the desired direction in the browsing history. This gesture (see Fig. 6), being the most popular (69%) within the sample, has been chosen for inclusion in the final dictionary.

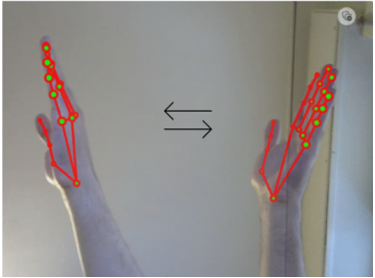

Figure 6: **browsing history gesture.**

The movement for navigating in history should be completed within a timeframe of *t ≤ 2s*. This timing is justified by the shared use of the open hand gesture for various other actions, as will be discussed later. Establishing these time limits is essential to ensure accurate identification of each action by the system.

## 4.6 Panning in resized page (Zoom >100%)

Only 5 groups (31%) out of 16 choose to include the action of panning in a resized page when the zoom of the viewport is >100%. 3 groups (60%) use the following gesture sequence: beginning with an open hand closing where no buttons or links are present, then moving the hand will correspond to moving the webpage, followed by releasing the hand. 1 group (20%) uses the middle, index and thumb finger of the right hand dragging the page in the desired direction. 1 group (20%) decided to use the index and middle finger where their movement will correspond to dragging the page.

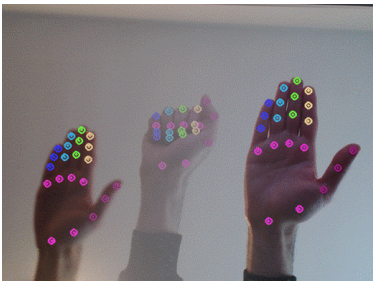

**Figure 7:** Panning gesture.

The gesture selected for panning within a resized page involves the closure of the hand in the whitespace, as illustrated in Figure 7. Recognized for its effectiveness and popularity among users, this gesture has been considered suitable for inclusion in the final dictionary.

## 4.7 Drag and Drop

Despite being implemented by only 5 groups, all of these groups (100%) chose a common gesture combination: closing the hand over the desired element and then moving the hand to the destination before releasing it (see Fig. 8). Although this gesture is the same as the one used for panning within a page, the system can distinguish between the two gestures. The drag-and-drop gesture includes a necessary condition that the hand must be closed on a webpage element. In contrast, for the resizing action, the hand must not be closed on buttons or links but only in the whitespace. This specification validates the inclusion of this combination of gestures in the final dictionary.

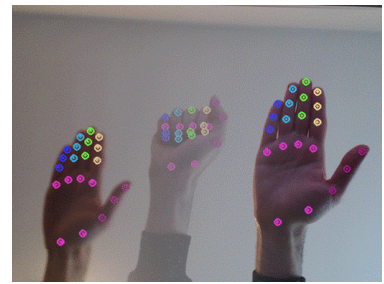

**Figure 8:** Drag and Drop gesture.

To enhance visual feedback, it is recommended to include the label of the element near the pointer, following the hand movement during dragging. This addition aims to make the movement of the element more evident and user-friendly. In the same way, for the previous panning action, an icon of a closed hand could be overlaid to indicate panning. This visual cue can suggest the panning action, providing users with a clear and intuitive understanding of the interaction. These visual enhancements can contribute to a more seamless and user-friendly touchless experience.

## 4.8 Interaction initialization and stop

Even though only 2 out of the 16 groups (13%) implemented the gesture to initiate the interaction, this task is considered crucial, particularly in a public context. Therefore, it will be included in the final dictionary. One group out of the two (50%) replicated the "greeting gesture" by moving the hand quickly from left to right. The other group (50%) proposed using the open hand, waiting for a duration of *t >= 3s* in front of the display to initialize the system. Both approaches contribute to the initiation of the system and provide users with options for starting the touchless interaction.

One of the group that implemented the gesture for initiating the interaction also introduced a corresponding gesture for ending the interaction. This gesture involves the use of an open hand held for a duration of *t >= 4s*, as shown in Figure 9. Following this gesture, a dialog window will appear, prompting the user to confirm the intention to terminate the interaction. This gesture serves a valuable purpose in addressing privacy concerns, as it clearly indicates the user's intent to conclude the interaction. This can be particularly crucial when users have entered (personal) information into the system and may need to physically move away from the area, leaving it unattended. Including such a gesture in the touchless interface adds an extra layer of security and user control over their interactions.

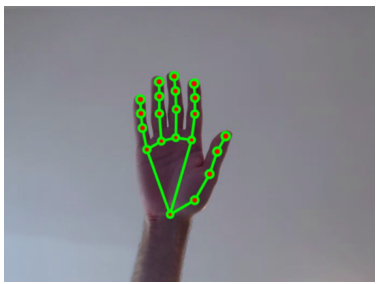

**Figure 9:** Initialization and stop gesture.

To start the interaction, in the final dictionary we incorporated the "greeting gesture", namely moving the hand quickly from left to right and vice versa. This is a spontaneous and shared gesture, also proposed in other touchless interaction [13].

To stop the interaction, users must hold a still hand for *t ≥ 4s*. This design allows the gesture to work seamlessly at any phase of web navigation, providing users with the flexibility to exit the interaction at their discretion. This comprehensive approach enhances user control and contributes to a more intuitive and user-friendly touchless interface.

## 4.9 Homepage access

Only 1 group out of 16 (6%) created a gesture to quickly return to the homepage. To execute it, both hands need to be raised and rotated by 90° so that both palms face each other, then quickly join them together. This gesture replicates the well known please gesture filling it with a further meaning, connoting reunion, and return. In addition, the two hands joined together form a triangular shape, which recalls the roof of a house. The gesture will take the user to the homepage.

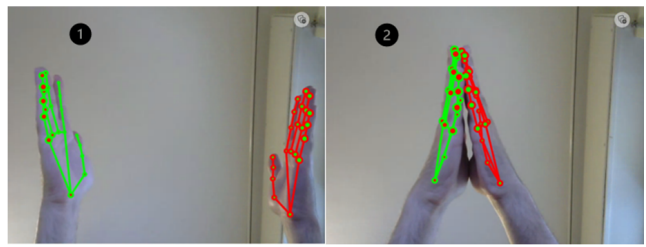

**Figure 10:** homepage access gesture.

This gesture is natural distinguishable from others, avoiding thus potential conflicts and being eligible for the final dictionary.

## 4.10 Page reload

Only one group (6%) out of the 16 introduced a gesture to refresh the page. The proposed gesture involves pointing the index finger towards the display and making a complete clockwise rotation of the index finger, forming a spiral, as shown in Fig. 11. The completion of the spiral triggers the web page to reload. This gesture has been considered eligible for inclusion in the final dictionary. It offers users an intuitive and distinguishing way to initiate the page refresh action during touchless interactions.

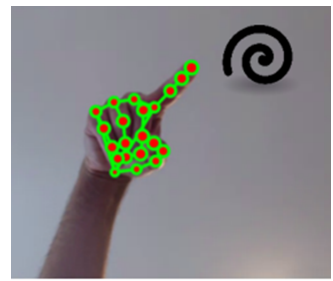

**Figure 11:** Page reload gesture.

The proposed gesture can indeed be useful in scenarios where users encounter errors that necessitate a page refresh or when they need to clear all entered data. If explained thoroughly during the tutorial phase, this gesture can be an easily remembered and quite intuitive. Providing users with a clear understanding of its function and benefits will contribute to a more effective and user-friendly experience.

## 4      Conclusion and future work

In this paper, we presented the results of an empirical evaluation of mid-air gestures, carried out with 56 students who acted as designers and 273 external participants, who were involved either as co-designers or as evaluators of the proposed gestures.

Our main contribution is the definition of a dictionary of mid-air gestures optimized for web browsing, and tested during an interaction with a university classroom search service.

In addition, however, our study allowed us to observe recurring behaviors which might be interesting also for designers of touchless interfaces who are approaching different domains. In particular, we found that users tend to adopt gestures primarily designed for touch-based and mouse-based interfaces, such as those we can find in most mobile devices, also for touchless interactions. The problem with this approach is that these gestures can be inefficient when they are used for more complex tasks in a touchless environment, leading to misinterpretation of user intent. One of the reasons which might have led the participants in our study to devise and/or positively assess such gestures is that laptops were used to simulate touchless devices. Consequently, all gestures were performed at a relatively small distance (approximately 50 cm) from the screen, an unlikely condition when interacting with large displays.

As future work, we deem it important to further evaluate the dictionary we have proposed. Firstly, we aim at testing the proposed gestures in a more realistic context, thus actually using large displays. Secondly, we need to verify that the adjustments and additions we have made to the original gestures identified by the designers (e.g., regarding timings) do not compromise their guessability.

A limitation in our study is that designers focused on a single specific context, i.e., interaction with a university classroom search service. Thus, the dictionary does not include actions which were not considered relevant for this scenario, either by the original designers or by the authors. Hence, it might not be possible to transpose the dictionary to a different web context without any additions and modifications.

Finally, we realize that, if large touchless displays were actually to be installed on the university campus, it would be advisable to make a tutorial for first-time users easily and always accessible, thus allowing them to practice mid-air gestures and gain familiarity with the dictionary. In fact, building confidence in the end user is crucial to minimizing potential errors in gesture enactment.

## ACKNOWLEDGMENTS

We thank all the students of the HCI 2022 course for their contributions to designing, evaluating, and enabling the progression of this research. Their efforts have played a significant role in advancing the understanding and development of our work in this field.